\documentclass[12pt]{article}
\usepackage{amsmath}
\usepackage{amssymb}\usepackage{epsf}
\usepackage{graphicx,color}
\usepackage{eepic}
\usepackage{latexsym}
\usepackage{epsfig}
\usepackage{cite}
\newcommand{\cO}{{\cal O}}
\newcommand{\be}{\begin{equation}}
\newcommand{\ee}{\end{equation}}

\begin{document}

\begin{center}
{\Large\bf  Calculations of ${\cal O}(p^6)$ Resonance Coupling
Constants in the Scalar Sector of
 the ENJL Model }
\\[10mm]
{\sc M.~X.~Su and L.~Y.~Xiao}
\\[2mm]
{\it  Department of Physics, Peking University, Beijing 100871,
P.~R.~China}
\\[5mm]
\today
\end{center}

\vskip 1cm
\begin{abstract}
We derive the scalar resonance coupling constants of resonance
chiral theory from the Extended Nambu Jona-Lasinio model by using
heat-kernel expansion.
\end{abstract}\vskip .5cm


Effective field theories(EFT) are useful tools in particle physics.
In the low-energy regime, the effective theory of hadrons is chiral
perturbation theory~\cite{chpt,U3-chpt}($\chi$PT). The explicit
degrees of freedom of $\chi$PT are the lightest pseudo-Goldstone
bosons from the spontaneous chiral symmetry breaking. The structure
of the effective lagrangian is determined by chiral symmetry and the
discrete symmetry of QCD. It is organized as an expansion in
derivatives of the Goldstone fields and in powers of the light
current quark mass ($m_q$).

The effective lagrangian depends on a number of low-energy
constants(LECs), which are not determined by symmetry, encoding the
underlying QCD dynamics. In principle, if one can solve QCD, we can
get all the coupling constants from QCD. Since in the low energy
regime, QCD is highly non-perturbative, it is hard to calculate
these coupling constants from QCD directly. \footnote{Attempts have
been made in expressing the low energy constants in terms of QCD
operators~\cite{WangQ}.} On the other side, how to understand ``high
energy physics'' from the low energy theory is also a very
interesting problem. In the original, resonance chiral theory
lagrangian(R$\chi$T)~\cite{rcht} was proposed as the most general
chiral invariant lagrangian that contributed to ${\cal O}(p^4)$
LECs. Therefore, LECs can be estimated by the masses and couplings
of the heavier resonances.

Progress has also been made along the direction by using $1/N_C$
expansion~\cite{NC1,NC2,NC3} and dispersion theory techniques
recently in Ref.~\cite{XiaoGuo}. More recent works have already
applied large-$N_C$ techniques to estimate subsets of the ${\cal
O}(p^6)$ LECs\cite{Moussallam:1997xx,
Knecht:2001xc,Ruiz-Femenia:2003hm,
Bijnens:2003rc,Cirigliano:2004ue,Cirigliano:2005xn,rcht-op6}. In
Ref.~\cite{rcht-op6}, the authors developed a systematic method to
calculate the contributions of resonances to the full ${\cal
O}(p^6)$ $\chi$PT to leading order in $1/N_C$.

However, another question arises naturally: how we determine the
resonance couplings? In the original R$\chi$T paper~\cite{rcht}, the
authors studied the resonances couplings phenomenologically to
determine the LECs from different physical processes. When we
analyze the ${\cal O}(p^6)$ LECs \cite{ChPT-op6}, the number of
couplings is increased significantly. Even though large $N_C$
technique and OPE matching conditions can be applied to give some
constraints of couplings as done in Ref.~\cite{rcht-op6}, it is
still insufficient to determine all the couplings. Therefore, it is
meaningful to calculate all these couplings via some QCD-inspired
models as a guide for future research.

In this letter, the extended Nambu Jona-Lasinio
model(ENJL)~\cite{NJL} is treated as an approximate model of QCD and
all the scalar resonances couplings ($\lambda_i^{S}$,
$\lambda_i^{SS}$ and $\lambda_i^{SSS}$) in R$\chi$T are derived
using heat-kernel method~\cite{heatkernel}.

The resonance chiral lagrangian can be organized in the following
way:
\begin{eqnarray}\label{eq:lag1}
 {\cal L}_\infty \ &=&  {\cal L}^{\rm GB}_{(2)}
 \ + \  {\cal L}^{\rm GB}_{(4)} \ + {\cal L}^{\rm GB}_{(6)}
\nonumber\\ &&  + \ {\cal L}^{R}_{\rm kin}  \ + \ {\cal L}^{R}_{(2)}
\ + \ {\cal L}^{R}_{(4)} \ + \ {\cal L}^{ R R}_{(2)} \ + \ {\cal
L}^{ R R R}_{(0)}  \ ,
\end{eqnarray}
where ${\cal L}^{\rm GB}_{(n)}$ is the Goldstone chiral Lagrangian
of $\cO(p^{n})$, ${\cal L}_{\rm kin}^{R}$ is the resonance kinetic
term, and ${\cal L}^{[...]}_{(n)} $ is a sum of monomials involving
the number of resonances specified in $[...]$ with chiral building
blocks of order $p^n$. The main approximations involve truncating
the hadronic spectrum to a finite number of states.
Usually\cite{rcht,rcht-op6}, the lowest lying resonance multiplets
with given $J^{PC}$ are considered.

The lagrangian and the resonance couplings of ${\cal L}^R_{(2)}$ are
well studied in Ref.~\cite{rcht}, but not many researches have been
done relate to ${\cal L}^R_{(4)}$, ${\cal L}^{RR}_{(2)}$, ${\cal
L}^{RRR}_{(0)}$.

There are 70 independent operators for the lagrangian density linear
in resonance fields:
\begin{equation} \label{L4R}
{\cal L}_{(4)}^{R} = \sum_{i=1}^{22} \, \lambda_i^{V} \, {\cal
O}^V_i + \sum_{i=1}^{17} \, \lambda_i^{A} \, {\cal O}^A_i +
\sum_{i=1}^{18} \, \lambda_i^{S} \, {\cal O}^S_i + \sum_{i=1}^{13}
\, \lambda_i^{P} \, {\cal O}^P_i \ .
\end{equation}

For the lagrangian quadratic in the resonance fields there are a
total of 38 operators:
\begin{equation} \label{eq:l2rr}
{\cal L}_{(2)}^{RR} = \sum_{(i j) n} \, \lambda_n^{R_i R_j} \ {\cal
O}^{R_i R_j}_n  \ ,
\end{equation}
with $R_i R_j= VV,AA,SS,PP,SA,SP,SV,PV,PA,VA$.

For the lagrangian cubic in resonance fields there are 7 independent
operators:
\begin{equation} \label{eq:l0rrr}
{\cal L}_{(0)}^{RRR} = \sum_{( i j k )}  \, \lambda^{R_i R_j R_k } \
{\cal O}^{R_i R_j R_k } \ .
\end{equation}

All the corresponding monomials ${\cal O}_i^{R}$, ${\cal O}_i^{RR}$,
${\cal O}_i^{RRR}$ can be found in Ref.~\cite{rcht-op6}. We just
list the operators relate to our calculations. Operators ${\cal
O}_i^{S}$ are in table.~\ref{linear}, ${\cal O}_i^{SS}$ and ${\cal
O}_i^{SSS}$ are listed below:
\begin{table}[!h]
\begin{center}
\renewcommand{\arraystretch}{1.5}
\begin{tabular}{|c|c|||c|c|}
\hline \multicolumn{1}{|c|}{$i$} & \multicolumn{1}{|c|||}{Operator
${\cal O}^S_i$} & \multicolumn{1}{|c|}{$i$} &
\multicolumn{1}{|c|}{Operator ${\cal O}^S_i$}  \\
\hline \hline
 $\lambda_1$ & $ \langle S \, u_{\mu} u^{\mu} \, u_{\nu} u^{\nu}  \,
 \rangle $
&$\lambda_{10}$ & $i \,  \langle \, S \, \{ \, f_{+}^{\mu \nu} \, ,
\,
u_{\mu} u_{\nu} \, \}  \, \rangle $ \\
\hline
$\lambda_{2}$ & $ \langle \, S \, u_{\mu} \, u_{\nu} u^{\nu} \,
u^{\mu} \, \rangle $ &$\lambda_{11}$ & $i \, \langle \, S \, u_{\mu}
\, f_{+}^{\mu \nu} \, u_{\nu}
 \, \rangle  $\\
\hline
$\lambda_{2}$ & $ \langle \, S \, u_{\mu} u_{\nu} u^{\mu} u^{\nu} \,
\rangle $
 & $\lambda_{12}$ & $ \langle \, S \, \{ \, \nabla_{\alpha} \,
f_{-}^{\mu \alpha} \, , \,
 u_{\mu} \, \} \, \rangle $\\
\hline
$\lambda_{4}$ & $ i \, \langle \, S \, u^{\mu} \, \rangle \, \langle
\, \nabla_{\mu}  \,\chi_- \, \rangle $
& $\lambda_{13}$ & $ \langle \, S \, \chi_{+} \, \chi_{+} \, \rangle $\\
\hline
$\lambda_{5}$ & $\langle \, S \, \chi_- \, \rangle \, \langle
\,\chi_- \, \rangle$
& $\lambda_{14}$ & $ \langle \, S \, \chi_{-} \, \chi_{-} \, \rangle $ \\
\hline
$\lambda_{6}$ & $  \langle \, S \, \{ \, \chi_{+} \, , \, u^{\mu}
u_{\mu} \, \} \, \rangle $
&$\lambda_{15}$ & $\langle \, S \, f_{+ \mu \nu} \, f_{+}^{\mu \nu} \, \rangle $\\
\hline
$\lambda_{7}$ & $ \langle \, S \, u_{\mu} \, \chi_{+} \, u^{\mu} \,
\rangle $
&$\lambda_{16}$& $\langle \, S \, f_{- \mu \nu} \, f_{-}^{\mu \nu} \, \rangle $ \\
\hline
$\lambda_{8}$ & $i \, \langle \, S \, \{ \, u^{\mu} \, , \,
\nabla_{\mu} \, \chi_{-} \, \}
 \, \rangle $
& $\lambda_{17}$ & $\left\langle  \, S \, \nabla_{\alpha}
\nabla^{\alpha} \, \left( u_{\mu} \,
u^{\mu} \right) \, \right\rangle $ \\
\hline
$\lambda_{9}$ & $ \langle \, S \, \rangle \, \langle \, \chi_- \,
\rangle \, \langle \, \chi_- \, \rangle $ &$\lambda_{18}$ & $
\langle \, S \, \nabla_{\mu} \nabla^{\mu} \,\chi_{+} \,
\rangle $ \\
\hline \multicolumn{4}{c}{}
\end{tabular}
\caption{Operator ${\cal O}_i^{S}$.}\label{linear}
\end{center}
\vspace*{-1.cm}
\end{table}

$${\cal O}^{SS}:<SSu^\mu u_\mu>,<Su^\mu Su_\mu>,<SS\chi_+>,$$\\
$${\cal O}^{SSS}:<SSS>,$$
with $<\ldots>$ short for the trace in the flavor matrix space.

The ENJL four quark interactions
\begin{eqnarray}
\label{ENJL}
{\cal L}_{\rm ENJL}&=&\sum_i\bar q(i\partial\sl-{\cal M})q + {\cal
L}_{\rm NJL}^{\rm S,P} + {\cal L}_{\rm
NJL}^{\rm V,A} + {\cal O}\left(1/\Lambda_\chi^4\right),\\
{\rm with}\hspace*{1.5cm} {\cal L}_{\rm NJL}^{\rm S,P}&=& \frac{
8\pi^2 G_S }{ N_C \Lambda_\chi^2} \, { \sum_{i,j}}
\left(\overline{q}^i_R
q^j_L\right) \left(\overline{q}^j_L q^i_R\right) \nonumber\\
{\rm and}\hspace*{1.5cm} {\cal L}_{\rm NJL}^{\rm V,A}&=& -\frac{
8\pi^2 G_V}{ N_C \Lambda_\chi^2}\, { \sum_{i,j}} \left[
\left(\overline{q}^i_L \gamma^\mu q^j_L\right)
\left(\overline{q}^j_L \gamma_\mu q^i_L\right) + \left( L
\rightarrow R \right) \right] \ ,\nonumber
\end{eqnarray}
where $i,j$ are flavor indices, $N_C$ is the number of the colors,
$\Psi_{R,L} \equiv (1/2) \left(1 \pm \gamma_5\right) \Psi$ and the
couplings $G_S$ and $G_V$ are dimensionless quantities. We adopt the
same symbols and definitions as in Ref.~\cite{Bijnens}. The low
energy effective action of the ENJL model, at intermediate energies
below or of the order of a cut-off scale $\Lambda_\chi$, is expected
to be a reasonable effective realization of the standard QCD
lagrangian.

 In Ref.~\cite{Bijnens}, an effective lagrangian for mesons is
 derived from ENJL model by using heat-kernel
 expansion. The ${\cal O}(p^4)$ LECs and the coupling constants of
${\cal L}^{R}_{(2)}$ can be easily read from the effective
lagrangian. One can also use other regularization
method~\cite{Wuyueliang} to get similar effective lagrangian.
Following the same method as in Ref.~\cite{Bijnens} and taking
 into account the next leading order (${\cal O}(p^6)$) terms of
heat-kernel expansion, the coupling constants of ${\cal O}^{R}$,
${\cal O}^{RR}$, ${\cal O}^{RRR}$ can be given systematically. For
the ENJL model the vector and axial-vector are usually in the form
of Proca field but not in the antisymmetry tensor form. It is rather
complicated to calculate the vector and axial vector sector directly
from ENJL model, but for the scalar sector there is no such
obstruction.

 All the scalar resonances
couplings ($\lambda_i^{S}$, $\lambda_i^{SS}$ and $\lambda_i^{SSS}$)
are calculated here. We introduce $\Gamma_{n}$ denoting the
incomplete gamma function \be\label{gammafunction}
\Gamma(n-2,x=\frac{M_Q^2}{\Lambda_{\chi}^2}) = \int_{M_Q^2 /
\Lambda_{\chi}^2}^{\infty}{dz \over z}e^{-z} z^{n-2} ;
 \,\,\,\,\, n=1,2,3,...\ .
\ee The results are listed below, where we also keep higher order
terms in $1/\Lambda_\chi$ expansion,

\begin{eqnarray}
&&\lambda_{1}^{S} =\frac{N_{c}}{\left( 4\pi \right)
^{2}}\frac{\Gamma _{1}}{12\lambda_{S}{M_{Q}}}\left[
\left(2-7\frac{\Gamma _{2}}{\Gamma _{1}}+2\frac{\Gamma _{3}}{\Gamma
_{1}}\right)g_{A}^{4}-\left(2-\frac{\Gamma
_{2}}{\Gamma _{1}}\right)g_{A}^{2}\left( 1-g_{A}^{2}\right) \right],\\
&&\lambda_{2}^{S} =\frac{N_{c}}{\left( 4\pi \right)
^{2}}\frac{\Gamma _{1}}{12\lambda_{S}{M_{Q}}}\left[
\left(2-5\frac{\Gamma _{2}}{\Gamma _{1}}+2\frac{\Gamma _{3}}{\Gamma
_{1}}\right)g_{A}^{4}-\left(2-3\frac{\Gamma
_{2}}{\Gamma _{1}}\right)g_{A}^{2}\left( 1-g_{A}^{2}\right) +%
\left( 1-g_{A}^{2}\right) ^{2}\right] , \\
&&\lambda_{3}^{S} =\frac{N_{c}}{\left( 4\pi \right)
^{2}}\frac{\Gamma _{1}}{6\lambda_{S}{M_{Q}}}\left[
\left(2\frac{\Gamma
_{2}}{\Gamma_{1}}-\frac{\Gamma _{3}}{{\Gamma_{1}}}\right)g_{A}^{4}+\left(2-2\frac{\Gamma _{2}}{\Gamma_{1}}\right)%
g_{A}^{2}\left( 1-g_{A}^{2}\right) -\frac{1}{2}\left(
1-g_{A}^{2}\right) ^{2}\right],\\
 &&\lambda_{4}^{S}
=\frac{N_{c}}{\left( 4\pi \right) ^{2}}\frac{\Gamma_1}{3\lambda
_{S}N_{F}M_{Q}}\left(1-\frac{\Gamma _{2}}{2\Gamma
_{1}}\right)g_{A}^{2},
\\&&\lambda_{5}^{S}=\frac{N_{c}}{%
\left( 4\pi \right) ^{2}}\frac{ g_A}{4\lambda
_{S} N_F}\left[ \frac{\Gamma_0}{B}-%
\frac{2\Gamma _{1}}{3M_{Q}}g_{A}\right], \\
&&\lambda_{6}^{S} =\frac{N_{c}}{\left( 4\pi \right) ^{2}}\frac{\Gamma_0}{\lambda _{S}B}\left( \frac{1}{%
8}-\frac{5\Gamma _{1}}{6\Gamma_0}+\frac{\Gamma
_{2}}{3\Gamma_0}\right) g_{A}^{2},
\\&&\lambda_{7}^{S}=\frac{N_{c}}{\left( 4\pi \right) ^{2}}%
\frac{\Gamma_0}{\lambda _{S}B}\left( \frac{1}{4}-\frac{5\Gamma _{1}}{6\Gamma _{0}}+\frac{\Gamma _{2}}{3\Gamma _{0}}%
\right) g_{A}^{2}, \\
&&\lambda_{8}^{S} =\frac{N_{c}}{\left( 4\pi
\right) ^{2}}\frac{g_A}{2\lambda _{S}}\left[
-\frac{\Gamma_1}{6M_{Q}}\left(2-\frac{\Gamma
_{2}}{\Gamma _{1}}\right)g_{A}+\frac{\Gamma _{0}}{B}\left( 1-\frac{%
\Gamma _{1}}{\Gamma_0}\right)\right],
\\&&\lambda_{9}^{S}=\frac{%
N_{c}}{\left( 4\pi \right) ^{2}}\frac{\Gamma _{1}}{12\lambda _{S}N_{F}^{2}M_{Q}}g_{A}^{2}, \\
&&\lambda_{10}^{S} =\frac{N_{c}}{\left( 4\pi \right)
^{2}}\frac{\Gamma_1}{6\lambda _{S}M_Q}\left[ (-1+\frac{3\Gamma
_{2}}{2\Gamma _{1}})g_{A}^{2}+\left( 1-g_{A}^{2}\right) \right] ,\\
&&\lambda_{11}^{S}=\frac{N_{c}}{%
\left( 4\pi \right) ^{2}}\frac{\Gamma _{1}}{3\lambda_S M_{Q}}\left(1-\frac{\Gamma _{2}}{2\Gamma _{1}}\right)g_{A}^{2}, \\
&&\lambda_{12}^{S} =-\frac{N_{c}}{%
\left( 4\pi \right) ^{2}}\frac{\Gamma _{1}}{3\lambda_S M_{Q}}\left(1-\frac{\Gamma _{2}}{2\Gamma _{1}}\right)g_{A}, \\
&&\lambda_{13}^{S}=-\frac{N_{c}%
}{\left( 4\pi \right) ^{2}}\frac{3\Gamma _{0}M_{Q}}{4\lambda _{S}B^{2}}\left( 1-\frac{2\Gamma _{1}}{3\Gamma _{0}%
}\right) , \end{eqnarray}
\begin{eqnarray}
&&\lambda_{14}^{S} =\frac{N_{c}}{\left( 4\pi \right)
^{2}}\frac{1}{4\lambda
_{S}}\left( -\frac{\Gamma _{0}%
}{B}g_{A}+\frac{\Gamma _{1}}{3M_{Q}}g_{A}^{2}+\frac{\Gamma _{0}}{B^{2}}%
M_{Q}\right),
\\&&\lambda_{15}^{S}=\frac{N_{c}}{\left( 4\pi \right) ^{2}}\frac{\Gamma _{1}}{6\lambda
_{S}M_{Q}},
\\&&\lambda_{16}^{S}=0,
\\
&&\lambda_{17}^{S} =-\frac{N_{c}}{\left( 4\pi \right)
^{2}}\frac{\Gamma_1}{6\lambda _{S}M_{Q}}\left(2-\frac{\Gamma
_{2}}{\Gamma _{1}}\right)g_{A}^{2},
\\&&\lambda_{18}^{S}=-\frac{%
N_{c}}{\left( 4\pi \right) ^{2}}\frac{\Gamma _{0}}{2\lambda _{S}B}\left( 1-\frac{2\Gamma _{1}}{3\Gamma _{0}}%
\right),
\end{eqnarray}

\begin{eqnarray}
\lambda _{1}^{SS} &=&\frac{N_{c}}{\left( 4\pi \right) ^{2}}\frac{g_{A}^{2}}{%
\lambda _{S}^{2}}\left[ \frac{\Gamma _{0}}{2}-\frac{10\Gamma _{1}}{3}+\frac{%
4\Gamma _{2}}{3}\right] ,\\
\lambda _{2}^{SS} &=&\frac{N_{c}}{\left( 4\pi \right) ^{2}}\frac{g_{A}^{2}}{%
\lambda _{S}^{2}}\left[ \frac{\Gamma _{0}}{2}-\frac{5\Gamma _{1}}{3}+\frac{%
2\Gamma _{2}}{3}\right] ,\\
\lambda _{3}^{SS} &=&-\frac{N_{c}}{\left( 4\pi \right)
^{2}}\frac{1}{\lambda
_{S}^{2}}\frac{3}{B}\left( \Gamma _{0}-\frac{2\Gamma _{1}}{3}\right) ,\\
\lambda_0^{SSS}&=&-\frac{N_{c}}{\left( 4\pi \right) ^{2}}\frac{1}{%
\lambda _{S}^{3}}M_{Q}4\left( \Gamma _{0}-\frac{2\Gamma
_{1}}{3}\right),
\end{eqnarray}

where
\begin{eqnarray}
&&x=\frac{M_Q^2}{\Lambda_\chi^2},\\
&&\lambda_S^2 = {N_c \over 16\pi^2} {2 \over 3} \left[3\Gamma_0 -
2\Gamma_1 \right],\\
&&B=\frac {\Gamma_{-1}\cdot M_{Q}}{\Gamma_{0}\cdot g_{A}},
\end{eqnarray}
 $M_Q$ is the constituent quark mass and \begin{equation}
\label{gA} g_A = {1 \over 1 + 4 G_V x \Gamma_0}\ ,
\end{equation}
characterizing the $\pi$ -- $A_1$ mixing.

During the calculations, the equation of motion of the
pseudo-Goldstone from ${\cal O}(p^2)$ $\chi$PT , \be \nabla_\mu
u^\mu-\frac{i}{2}\left(\chi_--\frac{1}{N_F}<\chi_->\right)=0 \ee
with $N_F$ the number of the flavors, has been used to simplify our
final results.

The above calculation can be extended to vector and axial vector
sector by using field redefinition\cite{Bijnens:1995ii} and even the
total 90 ${\cal O}(p^6)$ low energy constants. It is worthwhile to
point out that this kind of calculations is model-dependent, and
ENJL model itself suffers for its own problem~\cite{Su07} in the
scalar sectors. However, our calculations afford a reference for
further researches in resonance chiral theory.

\vspace*{-0.5cm}
\section*{Acknowledgments}

We are indebted to Professor Han-Qing Zheng for useful comments.
During this work we
benefited from helpful discussions with 
Zhi-Hui Guo and
Juan Jose Sanz-Cillero.  This work is supported in part by National
 Nature Science Foundations of China under contract number
 10575002, 
 10421503.


\end{document}